\documentclass[aps,floatfix]{revtex4}

\usepackage[dvips]{graphicx,color}
\usepackage{amsmath,amssymb}

\newcommand{\bu}{{\bf u}}
\newcommand{\buC}{{\bf u_C}}
\newcommand{\bup}{{\bf u'}}

\newcommand{\bxp}{{\bf x'}}
\newcommand{\xp}{{x^\prime}}
\newcommand{\zp}{{z^\prime}}

\begin{document}
\title{Turbulent-laminar patterns in plane Couette flow}
\author{Dwight Barkley}
\affiliation{Mathematics Institute, University of Warwick, 
  Coventry CV4 7AL, United Kingdom}
\email{barkley@maths.warwick.ac.uk,   www.maths.warwick.ac.uk/~barkley}

\author{Laurette S.\ Tuckerman}
\affiliation{LIMSI-CNRS, BP 133, 91403 Orsay, France}
\email{laurette@limsi.fr, www.limsi.fr/Individu/laurette}

\begin{abstract}
Regular patterns of turbulent and laminar fluid motion arise in plane Couette
flow near the lowest Reynolds number for which turbulence can be sustained.
We study these patterns using an extension of the minimal flow unit approach
to simulations of channel flows pioneered by Jim\'enez and Moin.  In our case
computational domains are of minimal size in only two directions.  The third
direction is taken to be large.  Furthermore, the long direction can be tilted
at any prescribed angle to the streamwise direction.  We report on different
patterned states observed as a function of Reynolds number, imposed tilt, and
length of the long direction. We compare our findings to observations in large
aspect-ratio experiments.

\vspace*{2cm}

Published version appears as
D.~Barkley and L.S.~Tuckerman, Turbulent-laminar patterns in 
plane Couette flow,
in {\it IUTAM Symposium on Laminar Turbulent Transition and 
Finite Amplitude Solutions}, T.~Mullin and R.~Kerswell (eds), 
Springer, Dordrecht, pp.~107--127 (2005).

\end{abstract}

\maketitle

\section{INTRODUCTION}

\begin{figure} 
\center{\includegraphics[width=2.5in]{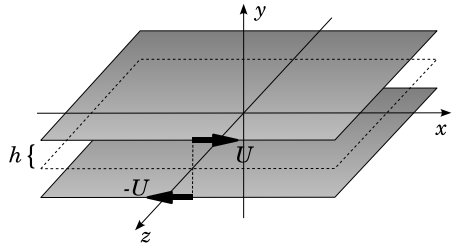}}
\caption{Plane Couette geometry.  Plates separated by a gap $2h$ move at
speeds $\pm U$.  The coordinate system we use is as shown, with $y=0$
corresponding to midgap. }
\label{fig:geometry}
\end{figure}

\noindent 
In this chapter we consider plane Couette flow -- the flow between two
infinite parallel plates moving in opposite directions.  This flow is
characterized by a single non-dimensional parameter, the Reynolds number,
defined as $Re= h U/\nu$, where $2h$ is the gap between the plates, $U$ is the
speed of the plates and $\nu$ is the kinematic viscosity of the fluid.  See
figure~\ref{fig:geometry}.  For all values of $Re$, laminar Couette flow
$\buC\equiv y{\bf\hat{x}}$ is a solution of the incompressible Navier-Stokes
equations satisfying no-slip boundary conditions at the moving plates.  This
solution is linearly stable at all values of $Re$.  Nevertheless it is not
unique.  In particular, for $Re$ greater than approximately $325$
\cite{Dauchot}, turbulent states are found in experiments and numerical
simulations.  Our interest is in the flow states found as one decreases $Re$
from developed turbulent flows to the lowest limit for which turbulence
exists.

\par
\begin{figure} 
\center{\includegraphics[width=12cm]{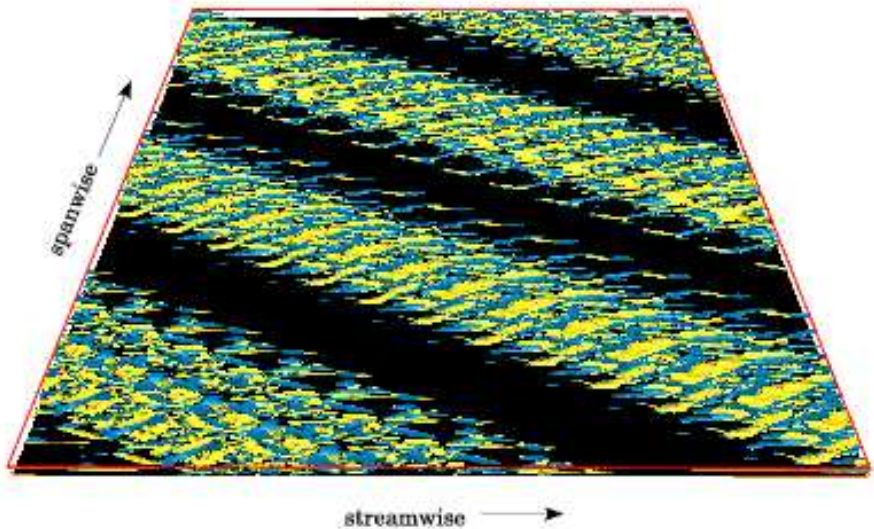} }
\vspace*{.5cm}
\caption{Turbulent-laminar pattern at Reynolds number 350.  Isosurfaces of
streamwise vorticity ($\omega=\pm0.5$) are shown at one instant in time.  For
clarity the bottom plate is shown in black while the top plate is
transparent. The streamwise and spanwise extent of the region shown are 60
times the plate separation $2h$.}
\label{fig:visualization}
\end{figure}

 Our work is motivated by the experimental studies of Prigent and
 coworkers (\citeyear{Prigent1},\citeyear{Prigent2},\citeyear{Prigent3},\citeyear{Prigent5})
on flow in a very large aspect-ratio
plane Couette apparatus. Near the minimum $Re$ for which turbulence is
sustained, they find remarkable, essentially steady, spatially-periodic 
patterns of turbulent and laminar flow. These patterns emerge
spontaneously from featureless turbulence as the Reynolds number is decreased.
Figure~\ref{fig:visualization} shows such a pattern from numerical
computations presented in this chapter.  Two very striking features of these
patterns are their large wavelength, compared with the gap between the plates,
and the fact that the patterns form at an angle to the streamwise direction.

\par Fluid flows exhibiting coexisting
turbulent and laminar regions have a significant
history in fluid dynamics.  In the mid 1960s 
a state known as spiral turbulence 
was first discovered \cite{Coles,vanAtta,ColesvanAtta} 
in counter-rotating Taylor--Couette flow.
This state consists
of a turbulent and a laminar region, each with a spiral shape.  The experiments
of Prigent {et al.}~
(\citeyear{Prigent1},\citeyear{Prigent2},
\citeyear{Prigent3},\citeyear{Prigent5})
in a very
large aspect-ratio Taylor--Couette system showed that in fact the turbulent and
laminar regions form a periodic pattern, of which the original observations of
Coles and van Atta comprised only one wavelength.  
\citeauthor{Cros} (\citeyear{Cros})
discovered large-scale turbulent spirals as well, in the shear
flow between a stationary and a rotating disk.  When converted to
comparable quantities, the Reynolds-number
thresholds, wavelengths, and angles are very similar for all of these
turbulent patterned flows. 

\section{METHODS}
\label{sec:methods}

\begin{figure} 
\center{\includegraphics[width=3.5in]{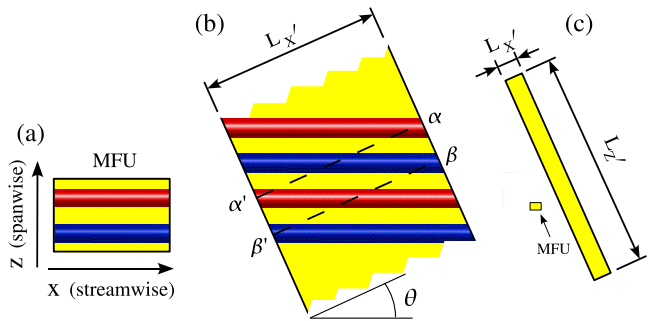}}
\caption{Simulation domains.  The wall-normal direction $y$ is not seen;
$L_y=2h$.  The bars represent streamwise vortex pairs with a spanwise
spacing of $4h$.  (The vortices are schematic; these are dynamic features of
the actual flow.)  (a) MFU domain of size $6h \times 4h$.  (b) Central portion
of a domain [on the same scale as (a)] tilted to the streamwise direction.
$\alpha$, $\alpha^\prime$ and $\beta$, $\beta^\prime$ are pairs of points
identified under periodic boundary conditions in $\xp$.  (c) Full tilted
domain with $L_\xp =10h$, $L_\zp = 120h$, $\theta=24^\circ$.  On this scale
the MFU domain, shown for comparison, is small.}
\label{fig:domains}
\end{figure}

\noindent Our computational technique \cite{BarkTuck05} 
extends the minimal flow unit methodology pioneered
by Jim\'enez and Moin~(\citeyear{Jimenez91}) and by
\citeauthor{Hamilton} (\citeyear{Hamilton}) and so we begin by
recalling this approach.  Turbulence near transition in plane Couette and
other channel flows is characterized by the cyclical generation and breakdown
of streaks by streamwise-oriented vortices.  The natural streak
spacing in the spanwise direction
is about 4--5$h$.  In the minimal flow unit approach, the smallest
laterally periodic domain is sought that can sustain this basic turbulent
cycle.  For plane Couette flow at $Re=400$, 
\citeauthor{Hamilton} (\citeyear{Hamilton}) 
determined this to be approximately $(L_x,L_y,L_z)=(4h,2h,6h)$. This 
domain is called the minimal flow unit (MFU).  The fundamental role of the
streaks and streamwise vortices is manifested by the fact that the spanwise
length of the MFU is near the natural spanwise streak spacing.
Figure~\ref{fig:domains}(a) shows the MFU in streamwise-spanwise coordinates.

We extend the MFU computations in two ways.  First we tilt the simulation
domain in the lateral plane at angle $\theta$ to the streamwise direction
[Figure~\ref{fig:domains}(b)]. We use $\xp$ and $\zp$ for the tilted
coordinates.  We impose periodic lateral boundary conditions 
on the tilted domain.  To respect
the spanwise streak spacing while imposing periodic boundary conditions in
$\xp$, the domain satisfies $L_\xp \sin \theta \simeq 4h$ for $\theta>0$.
(For $\theta=0$, we require $L_\xp \gtrsim 6h$.)  Secondly, we greatly extend
one of the dimensions, $L_\zp$, past the MFU requirement
[Figure~\ref{fig:domains}(c)], in practice between $30h$ and $220 h$, usually
$120 h$.

This approach presents two important advantages, one numerical and the other
physical.  First, it greatly reduces the computational expense of simulating
large length-scale turbulent-laminar flows.  Our tilted domains need only 
be long perpendicular to the turbulent bands. In the direction in which the
pattern is homogeneous, the domains are of minimal size, just large enough to
capture the streamwise vortices typical of shear turbulence. Second, the
approach allows us to impose or restrict the pattern orientation and 
wavelength.  We can thereby investigate these features and
establish minimal conditions necessary to produce these large-scale patterns.

We now present some further details of our simulations.  We consider the
incompressible Navier--Stokes equations written in the primed coordinate
systems. After nondimensionalizing by the plate speed $U$ and the half gap
$h$, these equations become
\begin{subequations} 
\begin{eqnarray} 
\label{eq:nse}
\frac{\partial \bup}{\partial t} + (\bup \cdot \nabla')\bup &=& 
  - \nabla' p' + \frac{1}{Re}\nabla'^2 \bup \quad \hbox{in $\Omega$},  \\
\label{eq:divcond}
\nabla' \cdot {\bup} &=& 0 \quad\hbox{in $\Omega$},
\end{eqnarray}
\end{subequations}
where $\bup(\bxp,t)$ is the velocity field and $p'(\bxp,t)$ is the static
pressure in the primed coordinate system, and $\nabla'$ is used to indicate
that derivatives are taken with respect to primed coordinates.  $\Omega$ is
the computational domain.  In these coordinates, the no-slip and
periodic boundary conditions are
\begin{subequations}
\begin{eqnarray} 
\label{eq:bcs}
\bup(x',y=\pm1,z') & = & \pm (\cos\theta, 0, \sin\theta) \\ 
\bup(x' + L_\xp,y,z') & = & \bup(x',y,z') \\
\bup(x',y,z' + L_\zp) & = & \bup(x',y,z')
\end{eqnarray}
\end{subequations}

\begin{figure} 
\center{\includegraphics[width=4in]{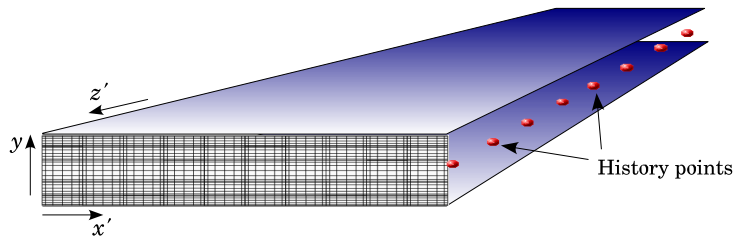}}
\caption{Simulation domain.  The $(\xp,y)$ grid is the actual spectral-element
mesh used for the case $L_{\xp}=10$.  
Only part of the $\zp$ direction is shown. 
In practice we use 32 history points in the $\zp$ direction.}
\label{fig:domain_3D}
\end{figure}

The equations are simulated using the spectral-element ($\xp$-$y$) -- Fourier
($\zp$) code {\tt Prism} \cite{Henderson}.  
We use a spatial resolution consistent with
previous studies \cite{Hamilton,Waleffe,WaleffePROC}. Specifically, for a domain with
dimensions $L_\xp$ and $L_y=2$, we use a computational grid with close to
$L_\xp$ elements in the $\xp$ direction and 5 elements in the $y$ direction.  
Within
each element, we usually use $6$th order polynomial expansions for the
primitive variables.  Figure \ref{fig:domain_3D} shows a spectral element mesh
used for the case of $L_\xp=10$.  In the $\zp$ direction, a Fourier
representation is used and the code is parallelized over the Fourier modes.
Our typical domain has $L_\zp=120$, which we discretize with
1024 Fourier modes or gridpoints. Thus the total spatial resolution
we use for the $L_\xp\times L_y \times L_\zp = 10\times 2\times 120$ domain
can be expressed as $N_\xp \times N_{y} \times N_\zp = 61 \times 31
\times 1024$.

We shall always use $(x,y,z)$ for the original streamwise, cross-channel,
spanwise coordinates (Figure~\ref{fig:geometry}).  We obtain usual streamwise,
and spanwise components of velocity and vorticity using
$u=u^\prime\cos\theta+w^\prime\sin\theta$ and
$w=u^\prime\sin\theta-w^\prime\cos\theta$, and similarly for vorticity.
The kinetic energy reported is the difference between the velocity $\bu$ and
simple Couette flow $\buC$, i.e.\ $E = \frac12((u-u_C)^2 + v^2 + w^2)$.

We have verified the accuracy of our simulations in small domains by comparing
to prior simulations \cite{Hamilton}. In large domains we have examined mean
velocities, Reynolds stresses, and correlations in a turbulent-laminar flow at
$Re=350$ and find that these reproduce experimental results from
Taylor--Couette \cite{ColesvanAtta} and plane Couette \linebreak
\cite{Hegseth96} flow.
While neither experimental study corresponds exactly to our case, the
agreement supports our claim that our simulations correctly
capture turbulent-laminar states.

The procedure we use to initiate turbulence is inspired by previous
investigations of plane Couette flow in a perturbed geometry.  We recall that
laminar plane Couette flow is linearly stable at all Reynolds numbers.  It has
been found, experimentally \cite{Bottin} and numerically \cite{BarkTuck99,TuckBark02},
that the presence of a wire \cite{Bottin} or a 
ribbon \cite{BarkTuck99,TuckBark02} oriented
along the spanwise direction causes the flow in the resulting geometry to
become linearly unstable to either a steady or a turbulent state containing
streamwise vortices.  We simulate such a flow with a ribbon which is
infinitesimal in the $\xp$ direction, occupies 30\% of the cross-channel
direction $y$ and spans the entire $\zp$ direction. At $Re=500$, the effect of
such a ribbon is to produce a turbulent flow quickly without the need to try
different initial conditions. Once the turbulent flow produced by the ribbon
is simulated for a few hundred time units, the ribbon can be removed and the
turbulence remains.  This procedure is used to initialize turbulent
states for the simulations to be described below.

\section{APPEARANCE OF TURBULENT-LAMINAR BANDS}

\subsection{Basic phenomenon}

\begin{figure} 
\center{\includegraphics[width=6cm]{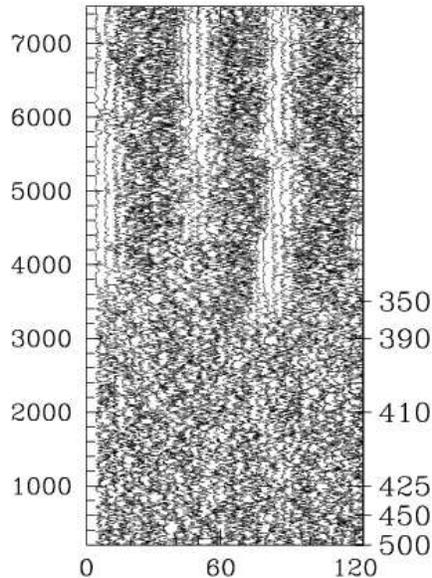} }
\caption{Space-time diagram. Kinetic energy $E(\xp=0,y=0,\zp)$ 
at 32 equally spaced points in $\zp$ in a domain with
$L_\xp\times L_y\times L_\zp = 10\times 2\times 120$ with
tilt $\theta=24^\circ$. The Reynolds number is decremented in
discrete steps (right). 
Three long-lasting and well-separated laminar regions emerge 
spontaneously from uniform turbulence as $Re$ is decreased.}
\label{fig:first_spacetime}
\end{figure}

\noindent We begin with one of our first simulations, in a domain tilted at angle
$\theta=24^\circ$.  This angle has been chosen to be close to that observed
experimentally near pattern onset.  The simulation shows the spontaneous
formation of a turbulent-laminar pattern as the Reynolds number is decreased.
We initiated a turbulent flow at $Re=500$ by perturbing laminar Couette flow
with a ribbon as described in Section~\ref{sec:methods}.  Time zero in
Figure~\ref{fig:first_spacetime} corresponds to the removal of the ribbon.
The flow is simulated for 500 time units at $Re=500$ and the
kinetic energy $E$ is measured at 32 points equally spaced in $\zp$ along the
line $\xp=y=0$ in the mid-channel shown in Figure~\ref{fig:domain_3D}. The
corresponding 32 time series are plotted at the corresponding values of $\zp$.
At $Re=500$, there is no persistent large-scale variation in the flow, a state
which we describe as uniform turbulence.  (This is not the homogeneous or
fully developed turbulence that exists at higher Reynolds numbers or in
domains without boundaries.)  At the end of 500 time units, $Re$ is abruptly
changed to $Re=450$ and the simulation continued for another 500 time
units. Then $Re$ is abruptly lowered to $Re=425$ and the simulation is
continued for 1000 time units, etc.~as labeled on the right in
Figure~\ref{fig:first_spacetime}.

At $Re=350$ we clearly see the spontaneous formation of a pattern. Out of
uniform turbulence emerge three regions of relatively laminar flow between
three regions of turbulent flow. (We will later discuss the degree to which the
flow is laminar.)  While the individual time traces are irregular, the pattern
is itself steady and has a clear wavelength of 40 in the $\zp$ direction.
This Reynolds number and wavelength are very close to what is seen
in the experiments.

\begin{figure} 
\begin{minipage}{7cm}
\center{\includegraphics[width=7cm]{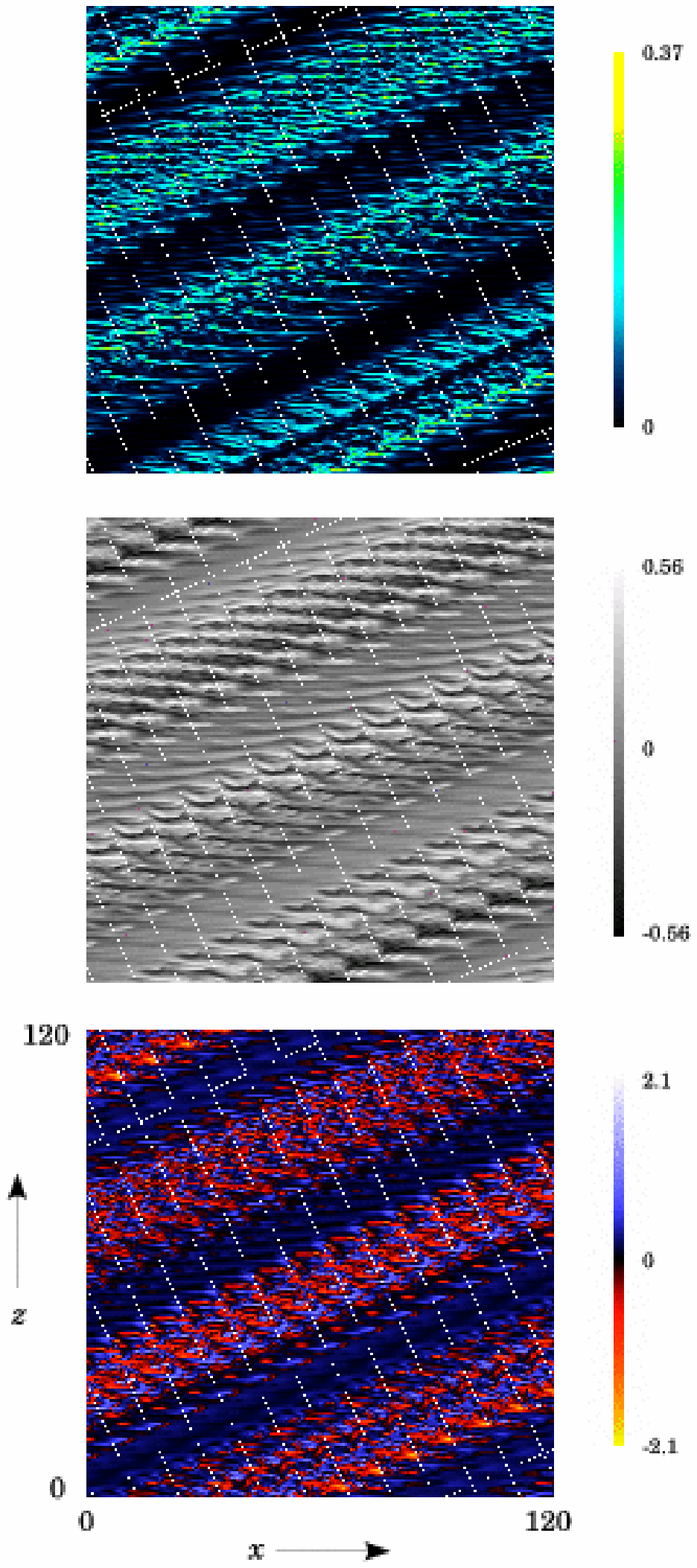} }
\caption{Turbulent-laminar pattern at $Re=350$.  The kinetic energy,
streamwise velocity, and streamwise vorticity are visualized in the $y=0$
plane, midway between and parallel to the moving plates.  
The computational domain
(outlined in white, tilted at angle $\theta=24^\circ$) is repeated
periodically to tile an extended region in $x$-$z$ coordinates.  Streamwise
streaks, with spanwise separation approximately $4h$, are visible at the edges
of the turbulent regions.}
\label{fig:three_boxes}
\end{minipage}
\hspace{0.5cm}
\begin{minipage}{4cm}
\center{\includegraphics[width=3cm]{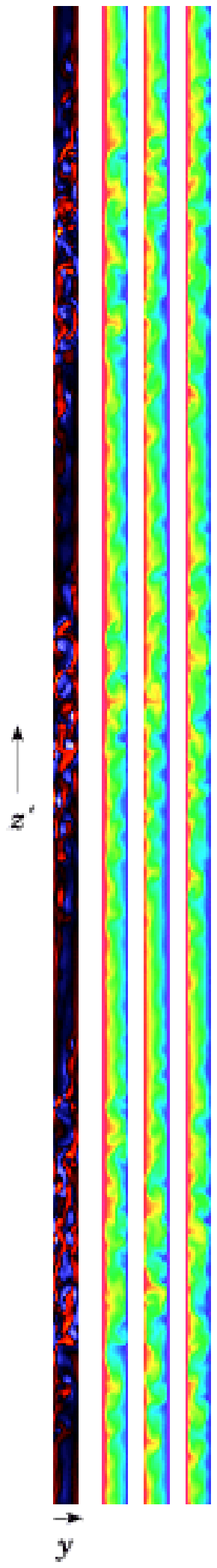} }
\caption{Turbulent-laminar pattern at $Re=350$ viewed between the moving
plates ($\xp=0$ plane). Left plot shows streamwise vorticity. The other three
plots show contours of streamwise velocity at three times separated by 100
time units (time increasing left to right).  The vorticity plot and the first
velocity plot correspond to the field seen in
Figure~\ref{fig:three_boxes}.}
\label{fig:four_cuts}
\end{minipage}
\end{figure}

\begin{figure} 
\center{ \includegraphics[width=4cm]{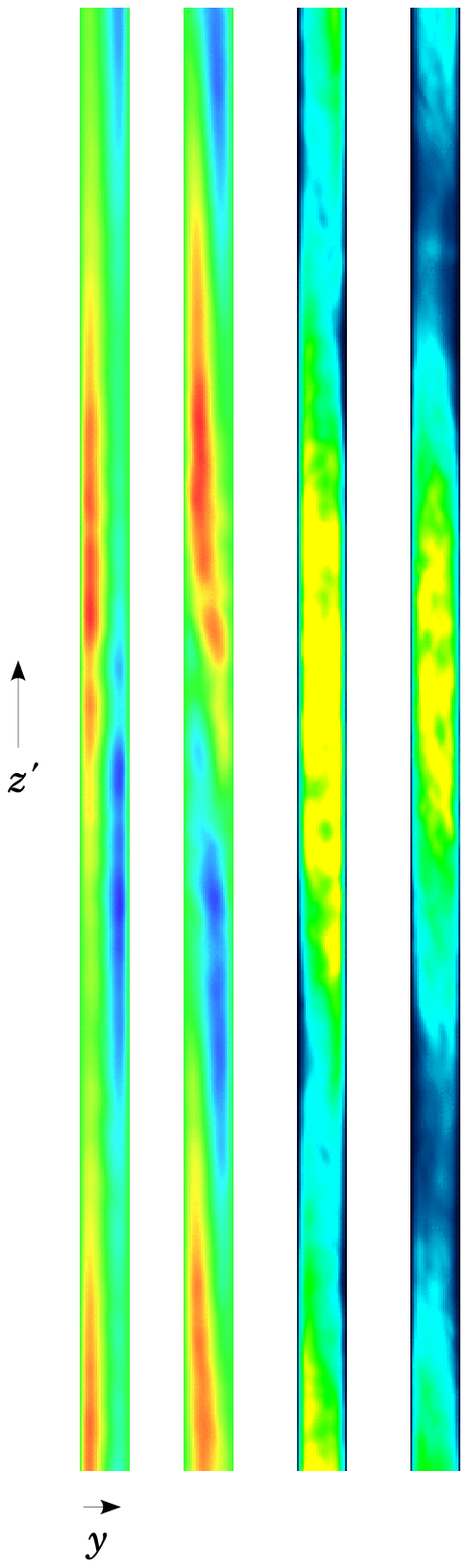} }
\caption{Mean and rms velocity fields for the turbulent-laminar pattern. 
From left to right: mean streamwise velocity, mean spanwise
velocity, rms streamwise velocity and rms spanwise velocity. The rms
velocities are maximal in the lightest regions. Only the central
half ($30 \le \zp \le 90$) of the computational domain is shown.}
\label{fig:mean_fluc}
\vspace*{1cm}
\center{ \includegraphics[width=10cm]{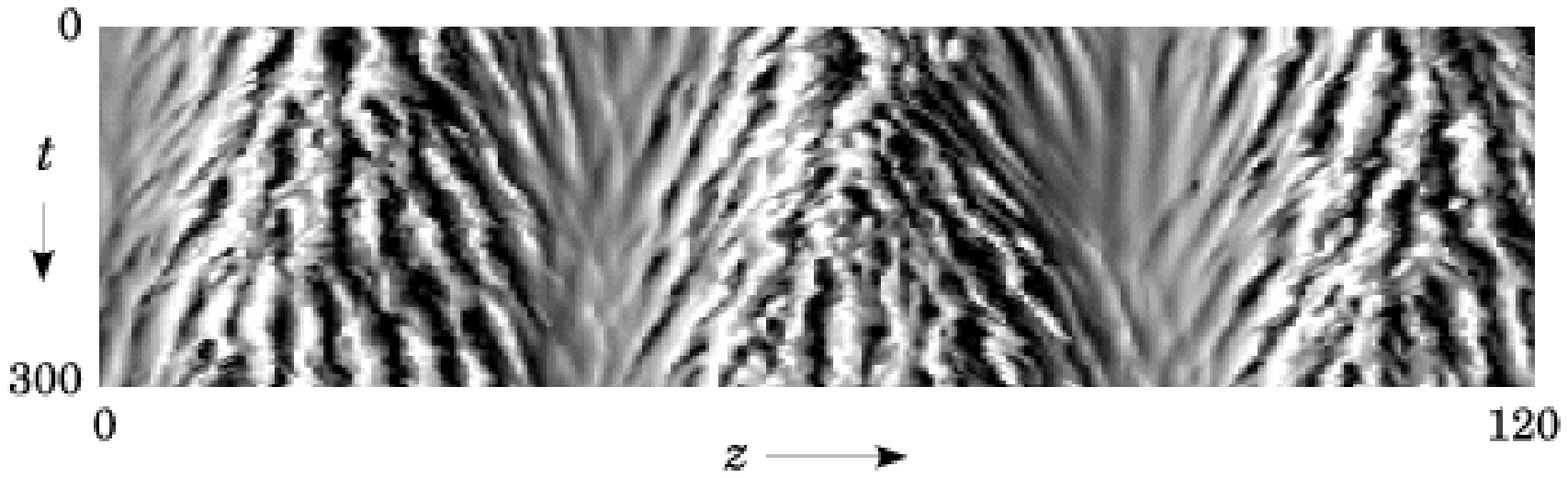} }
\caption{ Space-time plot showing dynamics of the turbulent-laminar pattern.
Streamwise velocity is sampled along a spanwise cut through the flow field
(the line $x=y=0$ in the reconstructed flows in Figure~\ref{fig:three_boxes}).
Time zero corresponds to the time of Figure~\ref{fig:three_boxes}.  
The streaks propagate away from the center of the turbulent regions toward the
laminar regions.  }
\label{fig:constant_x_spacetime}
\end{figure}

\subsection{Visualizations}

\noindent Figure~\ref{fig:three_boxes} shows visualizations of the flow at the final
time in Figure~\ref{fig:first_spacetime}.  Shown are the kinetic energy,
streamwise velocity, and streamwise vorticity in the midplane between the
plates. The computational domain is repeated periodically to tile an extended
region of the midplane.  The angle of the pattern is dictated by the imposed
tilt of the computational domain. The wavelength of the pattern is not imposed
by the computations other than that it must be commensurate with $L_\zp 120$.  The vorticity isosurfaces of this flow field were shown in
Figure~\ref{fig:visualization}.  Spanwise and cross-channel velocity
components show similar banded patterns.

Clearly visible in the center figure are streamwise streaks typical of shear
flows.  These streaks have a spanwise spacing on the scale of the plate
separation but have quite long streamwise extent.  We stress how these long
streaks are realized in our computations.  A streak seen in
Figure~\ref{fig:three_boxes} typically passes through several 
repetitions of the computational domain, as a consequence of the 
imposed periodic boundary conditions.  
In the single tilted rectangular computational domain,
a single long streak is actually computed as several adjacent
streaks connected via periodic boundary conditions.

Figure~\ref{fig:four_cuts} shows the streamwise vorticity and velocity fields
between the plates. The two leftmost images correspond to the same field 
as in Figure~\ref{fig:three_boxes}.  The
streamwise vorticity is well localized in the turbulent regions.  Mushrooms of
high- and low-speed fluid, corresponding to streamwise streaks, can be seen
in the turbulent regions of flow.  Dark velocity contours,
corresponding to fluid velocity approximately equal to that of the 
lower and upper moving
plates, are seen to reach into the center of the channel in the turbulent
regions.  In the center of the laminar regions, where the flow is relatively
quiescent (Figure~\ref{fig:first_spacetime}), there is very little streamwise
vorticity and the streamwise velocity profile is not far from that of laminar
Couette flow.  In particular, no high- or low-speed fluid reaches into the
center of the channel in these laminar regions.

In Figure~\ref{fig:mean_fluc} we show the mean and rms of the
streamwise and spanwise velocity components obtained from averages over
$T=2000$ time units.  
These results show that the mean flow is maximal at the boundaries
separating the turbulent and laminar regions while the fluctuations
are maximal in the middle of the turbulent bands.
This agrees with the experimental observations of 
of Prigent {et al.}~
(\citeyear{Prigent1},\citeyear{Prigent2},
\citeyear{Prigent3},\citeyear{Prigent5})
Note further that the regions of high fluctuation have approximately
the same rhombic shape as the turbulent regions shown by 
\citeauthor{ColesvanAtta} (\citeyear{ColesvanAtta})
in experiments on Taylor--Couette flow.
Finally, Figure~\ref{fig:constant_x_spacetime} shows a space-time plot 
of streamwise velocity along the spanwise line $x=y=0$.
Specifically, data is taken 
from reconstructed flows as in Figure~\ref{fig:three_boxes}.  Time zero in
Figure~\ref{fig:constant_x_spacetime} corresponds to the field in 
Figure~\ref{fig:three_boxes}.  Time is taken downward in this figure to allow for
comparison with a similar figure from the experimental study by Hegseth
(\citeyear{Hegseth96}: figure~6) showing the propagation of streaks away from the
center of turbulent regions.  Our results agree quantitatively with those of
Hegseth. We find propagation of streaks away from the center of the turbulent
regions with an average spanwise propagation speed of approximately 0.054 in
units of the plate speed $U$.  Translating from the diffusive time units used
by Hegseth, we estimate the average spanwise propagation speed of streaks in
his data to be approximately 0.060 at Reynolds number $420$.  This space-time
plot again shows the extent to which there is some small activity in the
regions we refer to as laminar.

\subsection{Average spectral coefficients}

\begin{figure}[t] 
\vspace*{-.5cm}
\includegraphics[width=12cm]{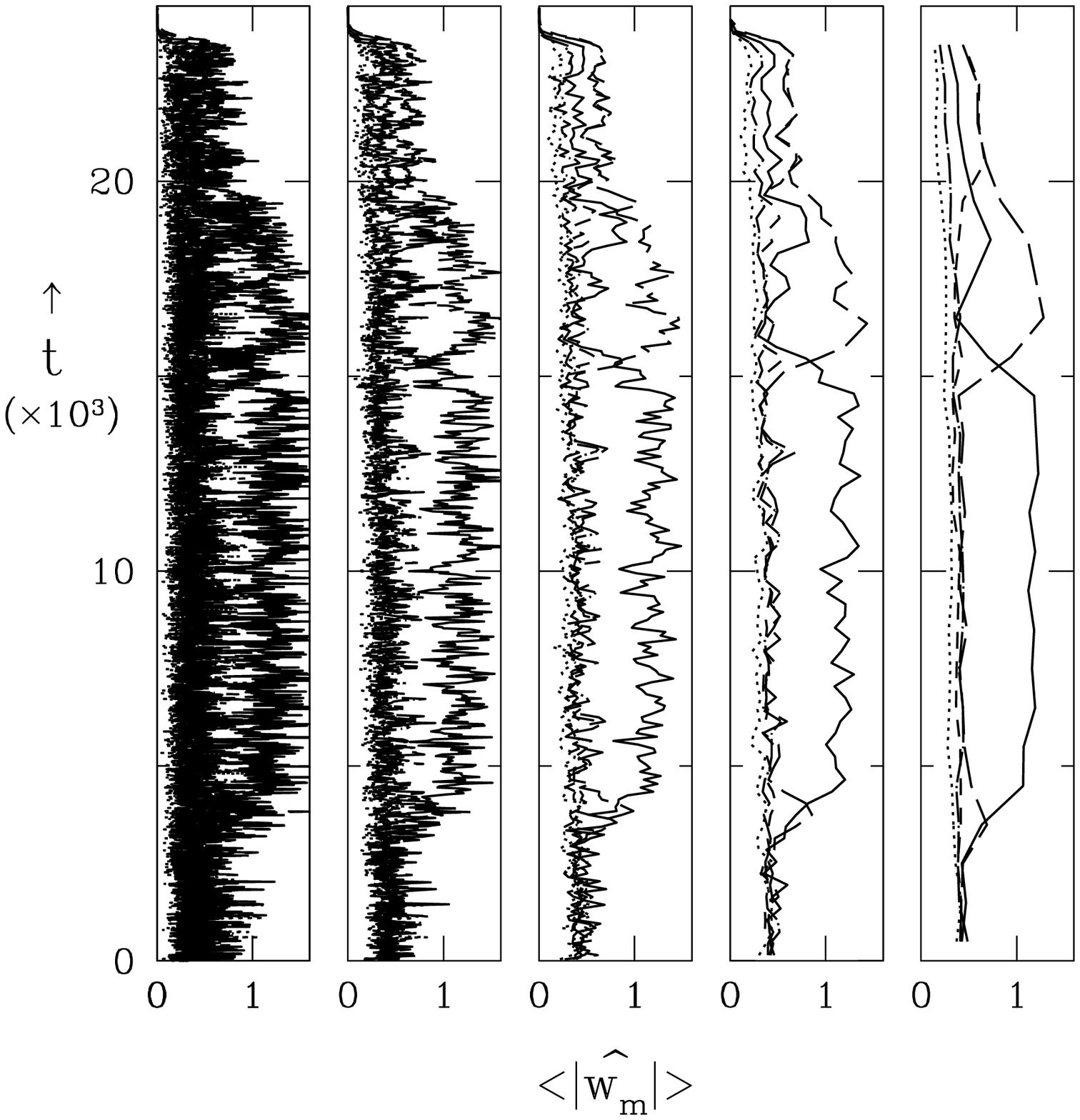}
\vspace*{-.5cm}
\caption{Evolution of $\langle |\hat{w}_m|\rangle$, which is an average over
time $T$ of the modulus of the Fourier transform in the $\zp$ direction
of 32 spanwise velocity samples taken along the line $(\xp=0,y=0)$.  The
components with wavenumber $m=3$ (solid curve), $m=2$ (long-dashed curve), 
$m=1$ (short-dashed curve) and $m=0$ (dotted curve) can be used as a
quantitative diagnostic of a turbulent-laminar pattern.  For example, the
dominance of the $m=3$ component
indicates a pattern containing three turbulent bands.
From left to right, the average is taken over $T=10$, $T=30$, $T=100$,
$T=300$, and $T=1000$.  }
\label{fig:spec_T_avg}
\end{figure}

\noindent We have determined a good quantitative diagnostic of the spatial periodicity
of a turbulent-laminar pattern.  We use the same data as that presented in
Figure~\ref{fig:first_spacetime}, i.e.~velocities at 32 points along the line
$\xp=y=0$ in the midplane along the long direction, at each interval of
$100$ time steps: 
$100\Delta t=1$.  We take a Fourier transform in $\zp$ of the spanwise
velocity $w$, yielding $\hat{w}_m$.  We take the modulus $|\hat{w}_m|$ to
eliminate the spatial phase.  Finally, we average over a time $T$ to obtain
$\langle |\hat{w}_m|\rangle$.  Figure \ref{fig:spec_T_avg} shows the evolution
of $\langle |\hat{w}_m|\rangle$ for wavenumbers $m=3$, $m=2$, $m=1$, and $m=0$
during one of our simulations (shown below in 
Figure~\ref{fig:spacetime_the24}, which is a continuation of that shown in 
Figure~\ref{fig:first_spacetime}).  As before, the vertical axis corresponds to time,
and also to Reynolds number, which was decreased in steps of $\Delta Re = 10$.
We average successively over $T=10$, $T=30$, $T=100$, $T=300$, 
and $T=1000$ and observe the
short-term fluctuations gradually disappear, leaving the long-term features
which will be discussed in the next section.  We have chosen $T=500$ as the
best compromise between smoothing and preserving the detailed evolution.

\section{DEPENDENCE ON REYNOLDS NUMBER}

\begin{figure}[t] 
\begin{minipage}{8cm}
\includegraphics[width=11.5cm]{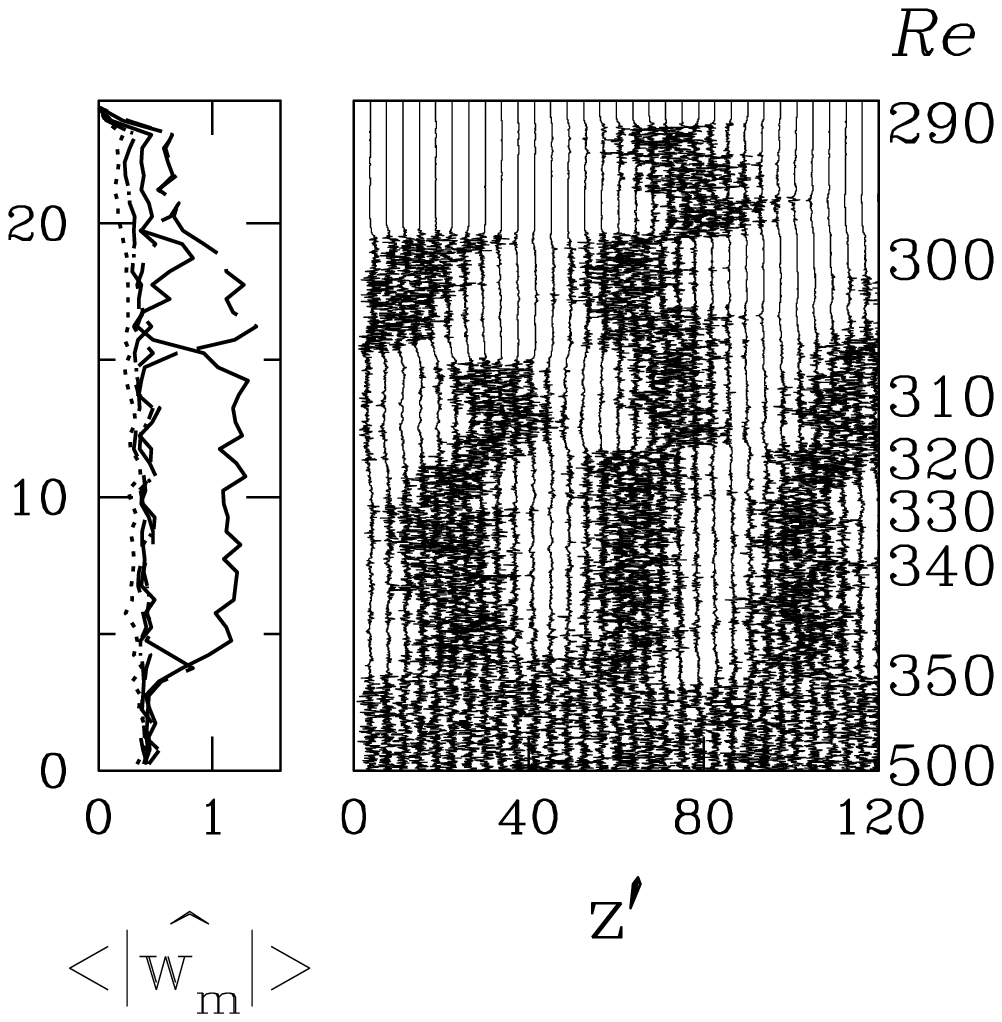}
\end{minipage}
\begin{minipage}{8cm}
\centerline{
\includegraphics[width=11.5cm]{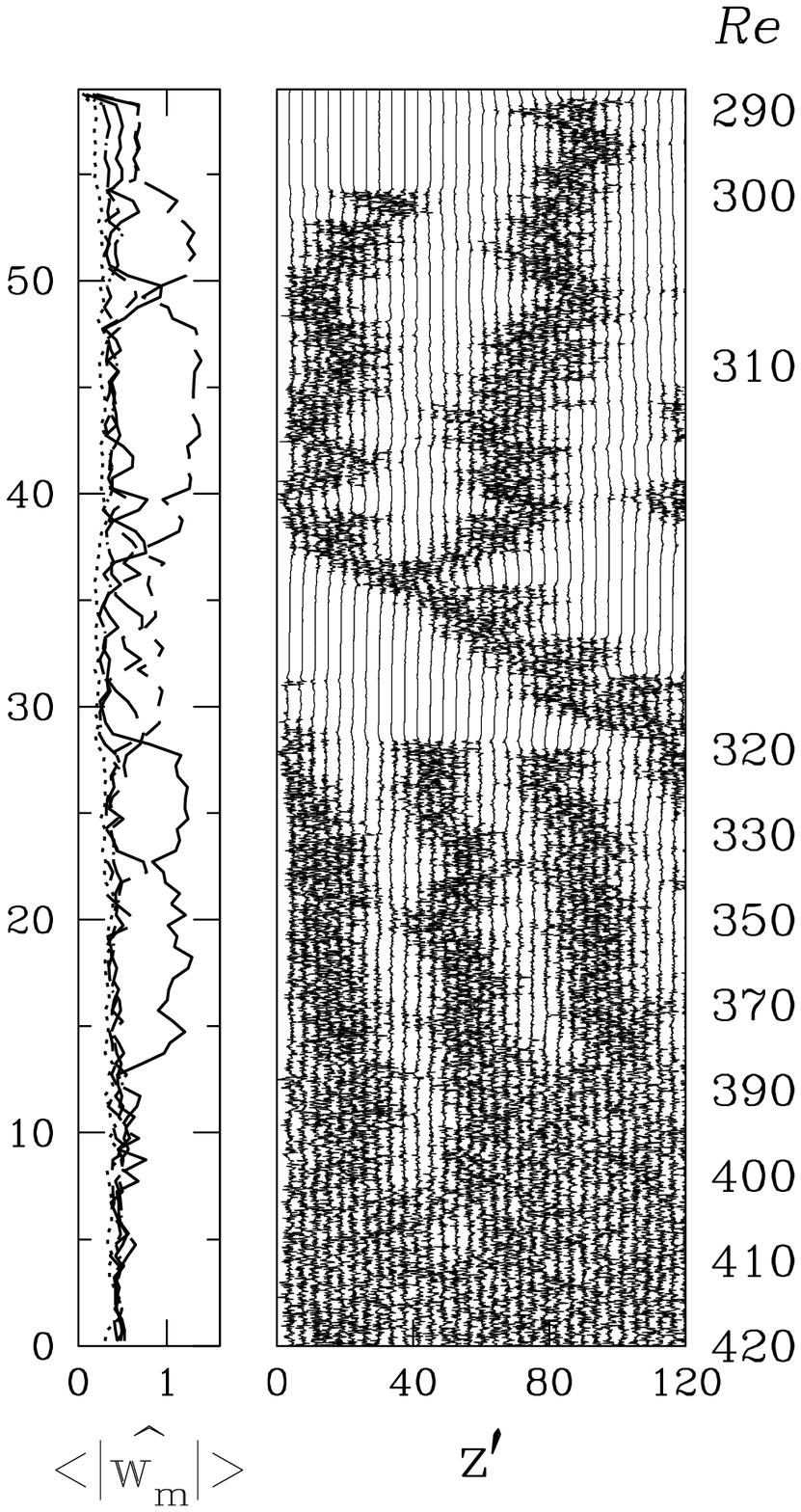}}
\end{minipage}
\caption{Two time series at $\theta=24^\circ$.  The Reynolds number is lowered
in steps, but at different instants to generate the two evolutions shown.  For
each case, we show $w$ at 32 points along the $\zp$ direction on the right and
the spectral components $\langle |\hat{w}_m|\rangle$ on the left.  Left:
uniform turbulence is succeeded by the formation of three bands, then two,
then a single band (a localized state) and finally by laminar Couette flow.
Right: two bands disappear almost simultaneously at $Re=320$.  The remaining
band moves toward the left, periodically emitting turbulent spurs, of which
one finally becomes a second turbulent band.}
\label{fig:spacetime_the24}
\end{figure}

\noindent We have investigated
in detail the Reynolds-number dependence of the $\theta=24^\circ$ case. To
this end, we have carried out two simulations, shown in Figure~\ref{fig:spacetime_the24}.  
In each the Reynolds number is lowered at discrete
intervals in time, but following a different sequence in the two cases.  For
each case, we present a space-time diagram of $E(\xp=0,y=0,\zp,t)$ at 32
values of $\zp$. The Reynolds-number sequence is shown on the right of each
diagram and the time (up to $T=59,000$) on the left.  Each space-time diagram
is accompanied by a plot showing the evolution of its average spectral
coefficients, as defined above. 

Careful observation of Figure~\ref{fig:first_spacetime} already shows a
laminar patch beginning to emerge at $Re=390$, consistent with experimental
observations: Prigent {et al.} (\citeyear{Prigent1},
\citeyear{Prigent2},\citeyear{Prigent3},\citeyear{Prigent5}) observed a
turbulent-laminar banded pattern with wavelength $46$ and angle $25^\circ$
when they decreased $Re$ below $Re=394$.
The space-time diagram on the left of Figure~\ref{fig:spacetime_the24}
shows a continuation of this simulation.
(Here, the Reynolds numbers intermediate between
500 and 350 are not shown to reduce crowding.)  
We see a sequence of
different states: uniform turbulence and the three-banded turbulent-laminar
pattern already seen are succeeded by a two-banded pattern (at $Re=310$), then
a state containing a single localized turbulent band (at $Re=300$), and
finally laminar Couette flow.  These features are reflected in the average
spectral coefficients.
The flow evolves from uniform turbulence (all components of about the same
amplitude) to intermittent turbulence, to a pattern containing three turbulent
bands (dominant $m=3$ component) and then two turbulent bands (dominant $m=2$
component), then a single band (dominant $m=1$ and $m=2$ components), and
finally becomes laminar (all components disappear).

In the simulation on the right, the Reynolds number is decreased more slowly.
A state with three bands appears at $Re=390$.  (Although a laminar patch
already appears at $Re=400$, it is regained by turbulence when $Re$ is
maintained longer at 400; this is not shown in the figure.)  Based on the
previous simulation shown on the left, we had expected the three turbulent
bands to persist through $Re=320$.  However here, instead, we see a rapid loss
of two bands, leaving only a single turbulent band.  This band moves to the
left with a well-defined velocity, emitting turbulent spurs toward the right
periodically in time.  Finally, after a time of $T=36000$, one of these spurs
succeeds in becoming a second turbulent band and the two bands persist without
much net motion. It would seem that the loss of the second band was premature,
and that at $Re=320$ one band is insufficient.
We then resumed the simulation on the left, maintaining $Re=320$ for
a longer time, and found that two bands resulted in this case as well.
Both simulations show two bands at $Re=310$, one band at $Re=300$, 
and laminar Couette flow at $Re=290$.

\subsection{Three states}

\begin{figure} 
\center{\includegraphics[width=10cm]{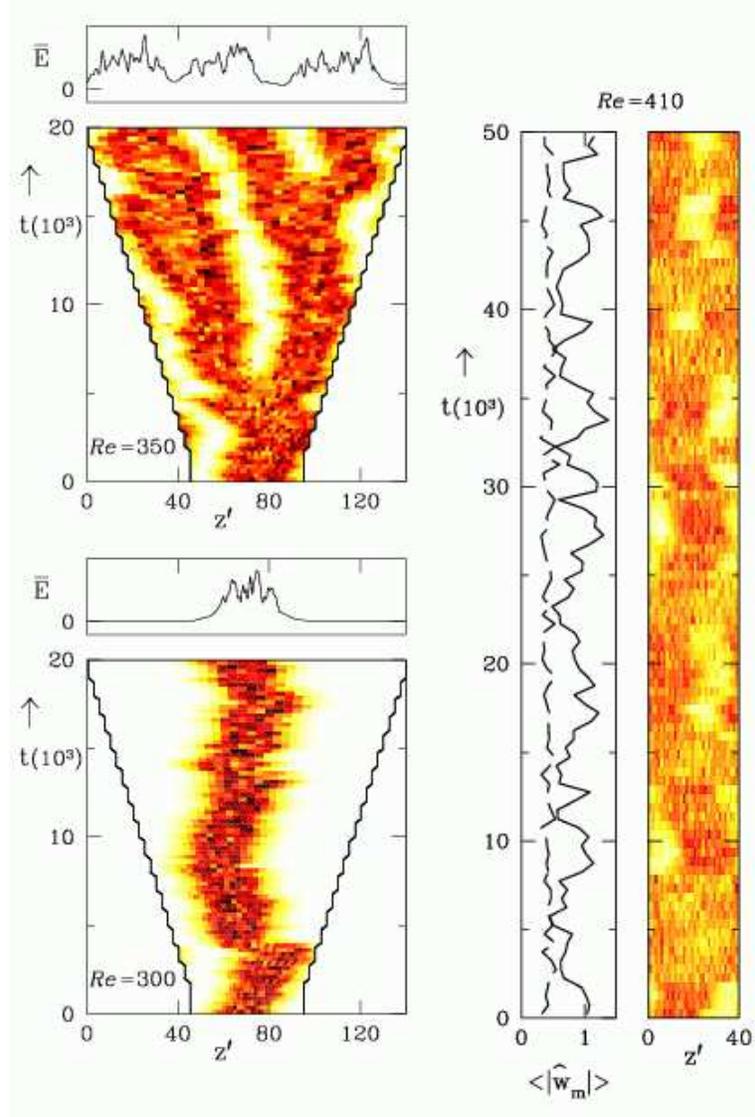} }
\caption{Simulations at $Re=350$, $Re=300$, and $Re=410$ illustrating
three qualitatively different regimes. For the simulations at $Re=350$
and $Re=300$, $L_\zp$ is increased from 50 to 140. The state at $Re=350$
is periodic: the turbulent band divides as $L_\zp$ is increased to retain a 
wavelength near 40. The final kinetic energy profile
$\bar{E}(\zp)$ is bounded away from zero. 
The state at $Re=300$ is localized: a single turbulent band persists, 
regardless of domain size and $\bar{E}(\zp)$ decays exponentially to zero
away from the band. The simulation at $Re=410$ is
carried out at $L_\zp=40$. The state is intermittent: laminar
regions appear and disappear and the average spectral coefficients
corresponding to $m=1$ (solid) and $m=0$ (dashed) oscillate erratically.}
\label{fig:three_states}
\end{figure}

\noindent The turbulent-laminar patterned states shown in 
Figure~\ref{fig:spacetime_the24} are of three qualitatively different types
\cite{BarkTuck05}.  We demonstrate this by carrying out three long
simulations, at three different Reynolds numbers, that are shown in 
Figure~\ref{fig:three_states}.  In this figure, the energy along the line
$\xp=y=0$ for the 32 points in $\zp$ has been averaged over windows of length
$T=500$ to yield a value shown by the shading of each space-time rectangle.

The simulations at $Re=350$ and $Re=300$ are carried out by increasing the
long direction of our domain, $L_\zp$, in regular discrete increments of 5
from $L_\zp=50$ to $L_\zp=140$. At $Re=350$, a single turbulent band is
seen when $L_\zp=50$.  This band divides into two when $L_\zp=65$ and a third
band appears when $L_\zp=130$: the periodic pattern adjusts to keep the
wavelength in the range 35--65. This is close to the wavelength
range observed experimentally by Prigent et al., which is 46--60.  
When the same protocol is followed at
$Re=300$, no additional turbulent bands appear as $L_\zp$ is increased.  We
call the state at $Re=300$ localized and note that turbulent spots are
reported near these values of $Re$ in the experiments.
The small $L_\xp$ of our computational
domain does not permit localization in the $\xp$ direction; instead localized
states must necessarily take the form of bands when visualized in the $x$-$z$
plane.

\sloppy The instantaneous integrated kinetic energy profile 
\[
\bar{E} \equiv \int\,d\xp\, dy \: E(\xp,y,\zp)
\]
is plotted at the final time for both cases.
For $Re=350$, $\bar{E}$ does not reach zero and the flow does not revert to
the simple Couette solution between the turbulent bands, as could also be seen
in the earlier visualizations (Figures \ref{fig:four_cuts},
\ref{fig:constant_x_spacetime}). In contrast, for $Re=300$, $\bar{E}$ decays to
zero exponentially, showing that the flow approaches the simple Couette
solution away from the turbulent band.  In this case, there is truly
coexistence between laminar and turbulent flow regions.

The simulation at $Re=410$ illustrates another type of behavior. In a domain
of length $L_\zp=40$, laminar or, rather, weakly-fluctuating regions appear
and disappear.  The spectral coefficients corresponding to 
$m=1$ (wavelength 40) and $m=0$ oscillate erratically.  
Similar states at similar Reynolds numbers are
reported experimentally by Prigent and coworkers
(\citeyear{Prigent1},\citeyear{Prigent2},\citeyear{Prigent3},\citeyear{Prigent5}), where they are
interpreted as resulting from noise-driven competition between banded patterns
at equal and opposite angles, a feature necessarily absent from our
simulations.

\section{DEPENDENCE ON ANGLE}

\subsection{Angle survey}

\begin{figure}[t] 
\centerline{\includegraphics[width=8cm]{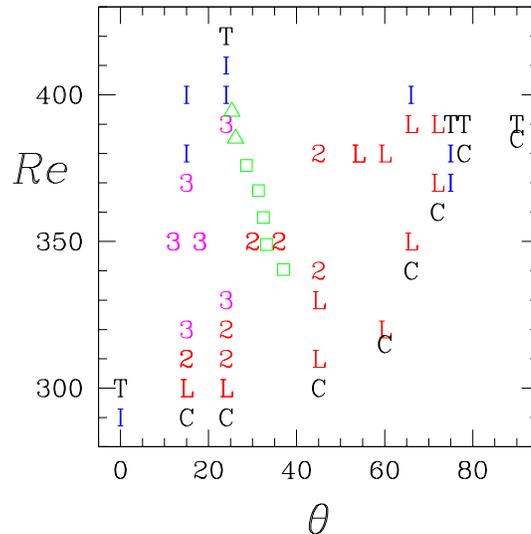}}
\vspace*{-.5cm}
\caption{ Patterns as a function of Reynolds number $Re$ and $\theta$, the
angle between the $L_\xp$ direction of our rectangular computational
domain and the  streamwise direction. The domain is
of size $L_\xp\times L_y \times L_\zp= (4/\sin\theta)\times 2 \times 120$.
For each angle, upper and lower limits in $Re$ are shown for each regime.
T: uniform turbulence (lower limit in $Re$). 
I: intermittent turbulence.  
3: pattern containing three turbulent-laminar bands, each of
approximate wavelength 40.
2: pattern containing two bands of approximate wavelength 60.
L: pattern containing one turbulent region, possibly localized.
C: laminar Couette flow (no patterns observed below this $Re$).
Open symbols show experimental observations of Prigent et al.~
Triangles: patterns with wavelength between 46 and 50.
Squares: patterns with wavelength between 50 and 60.}
\label{fig:scat_th_Re_trunc}
\end{figure}

\noindent We have explored the angles with respect to the streamwise direction at which
a turbulent-laminar pattern may exist.  The results are plotted in 
Figure~\ref{fig:scat_th_Re_trunc}.  We keep $L_{z\prime}=120$ and
$L_\xp=4/\sin\theta$.  The transition from uniform turbulence to
laminar Couette flow occurs via intermediate states which occupy a decreasing
range of $Re$ as $\theta$ is increased.  The sequence of states seen for
increasing $\theta$ at $Re=350$ is qualitatively the same as that
for decreasing $Re$ at $\theta=24^\circ$: uniform turbulence at
$\theta=0^\circ$, a turbulent-laminar pattern with three bands at
$\theta=15^\circ$ to $\theta=24^\circ$, two bands for $\theta=30^\circ$ and
$\theta=45^\circ$, a localized state for $\theta=66^\circ$, and laminar
Couette flow for $\theta^\circ \geq 72$.  
Thus far we have obtained patterns for angles between $15^\circ$ and 
$66^\circ$ and the number of bands decreases with angle.

Experimental data from Prigent and coworkers 
(\citeyear{Prigent1},\citeyear{Prigent3},\citeyear{Prigent5}) is also shown in 
Figure~\ref{fig:scat_th_Re_trunc}. The wavelengths, angles, and Reynolds
numbers reported ranged from $\lambda_\zp=46.3$ and $\theta=25.3^\circ$ 
at $Re=394$ to $\lambda_\zp=60.5$ and $\theta=37^\circ$ at $Re=340$.  
In these ranges of
angle and Reynolds number, we observe a similar trend,
since our wavelength (constrained here to be a divisor of 
$L_\zp=120$) increases from 40 to 60 as the number
of bands decreases from 3 to 2.
Between $Re=325$ and $Re=280$, experiments showed spots, 
which may correspond to some of the states we have labeled
as localized in Figure~\ref{fig:scat_th_Re_trunc}.
At present, we do not systematically distinguished localized
states from others containing one turbulent region but which
may not behave like Figure~\ref{fig:three_states}.
The threshold for intermittency is also difficult to define
and to determine.

The most striking difference between our computations
and the experimental data is that the range
of angles over which we find periodic turbulent-laminar patterns
(from $\theta=15^\circ$ to at least $\theta=45^\circ$)
is far greater than that seen in the experiment.
Patterns with angles outside of the experimental range
are likely to be unstable in a large domain in which
the angle is unconstrained.

Our computational technique requires that
the size of the domain be increased as $\theta$ 
decreases according to $L_\xp= 4/\sin\theta$ in order to 
respect the spanwise vortex or streak spacing; see Figure~\ref{fig:domains}. 
Hence the computational cost increases with decreasing $\theta$
and for this reason we have not as yet
investigated $\theta$ between $15^\circ$ and $0^\circ$.
For $\theta$ exactly $0^\circ$, this trigonometric constraint is lifted,
since the streamwise vortices and streaks would not extend
diagonally across the rectangular domain, but parallel
to its boundaries.
As $\theta$ increases, the domain size $L_\xp= 4/\sin\theta$ decreases,
as does the computation cost.
For $\theta$ between $45^\circ$ and $90^\circ$, for which $L_\xp$ is between
5.7 and 4, we reduce the number of spectral elements in the $\xp$ direction 
from 10 to 4 (see Figure~\ref{fig:domain_3D}).

\begin{figure}[t]
\vspace*{-2cm}
\centerline{\includegraphics[width=10cm]{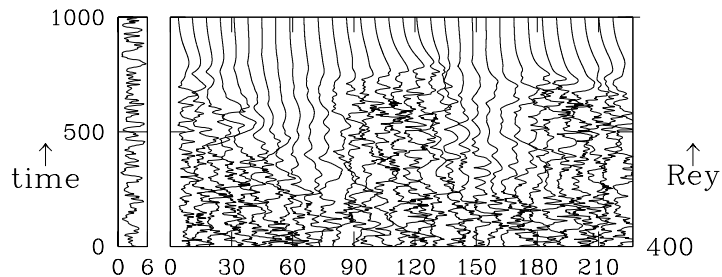}}
\vspace*{-3.5cm}
\caption{
Evolution for $\theta=90^\circ$.  
Two simulations at $Re=400$ are shown, in domains with $L_\xp=L_z=4$.
The domain shown on the right has $L_\zp=L_x=220$; that on the
left has $L_\zp=L_x=6$, close to the minimal flow unit 
(the two are not shown to scale).
The large domain supports a transient
pattern as an intermediate state between uniform turbulence and laminar Couette
flow, whereas the turbulence persists in the small domain.}
\label{fig:theta90}
\end{figure}
\subsection{Long streamwise direction}

\noindent For $\theta=90^\circ$, the domain has a long streamwise direction $L_\zp=L_x$
and a short spanwise direction $L_\xp=L_z$.  Figure~\ref{fig:scat_th_Re_trunc}
shows that, for $\theta=90^\circ$ and $L_x=120$, we obtain direct decay from
uniform turbulence to laminar Couette flow at $Re=385$.  We have varied $L_x$
and show the results in Figure~\ref{fig:theta90}.  When $L_x=220$, the
turbulence is extinguished at $Re=400$; a transient pattern of wavelength 110
can be seen.  But when $L_x=6$, we find that the turbulence persists down to a
value of $Re\approx 370$.  We recall that the minimal flow unit was proposed
by \citeauthor{Hamilton} (\citeyear{Hamilton})
as the smallest which can support the streak and
streamwise-vortex cycle and maintain turbulence; the flow becomes laminar when
either of the dimensions are reduced below their MFU values.  However, 
Figure~\ref{fig:theta90} shows that turbulence can also sometimes 
be extinguished by increasing
$L_x$.  Simulations in domains with a long streamwise and a short spanwise
dimension have also been carried out by Jim\'enez et al.~\cite{Jimenez05} with
the goal of understanding the role of the streamwise dimension, e.g.~streak
length.

\begin{figure}[t] 
\begin{minipage}{7cm}
\includegraphics[width=11cm]{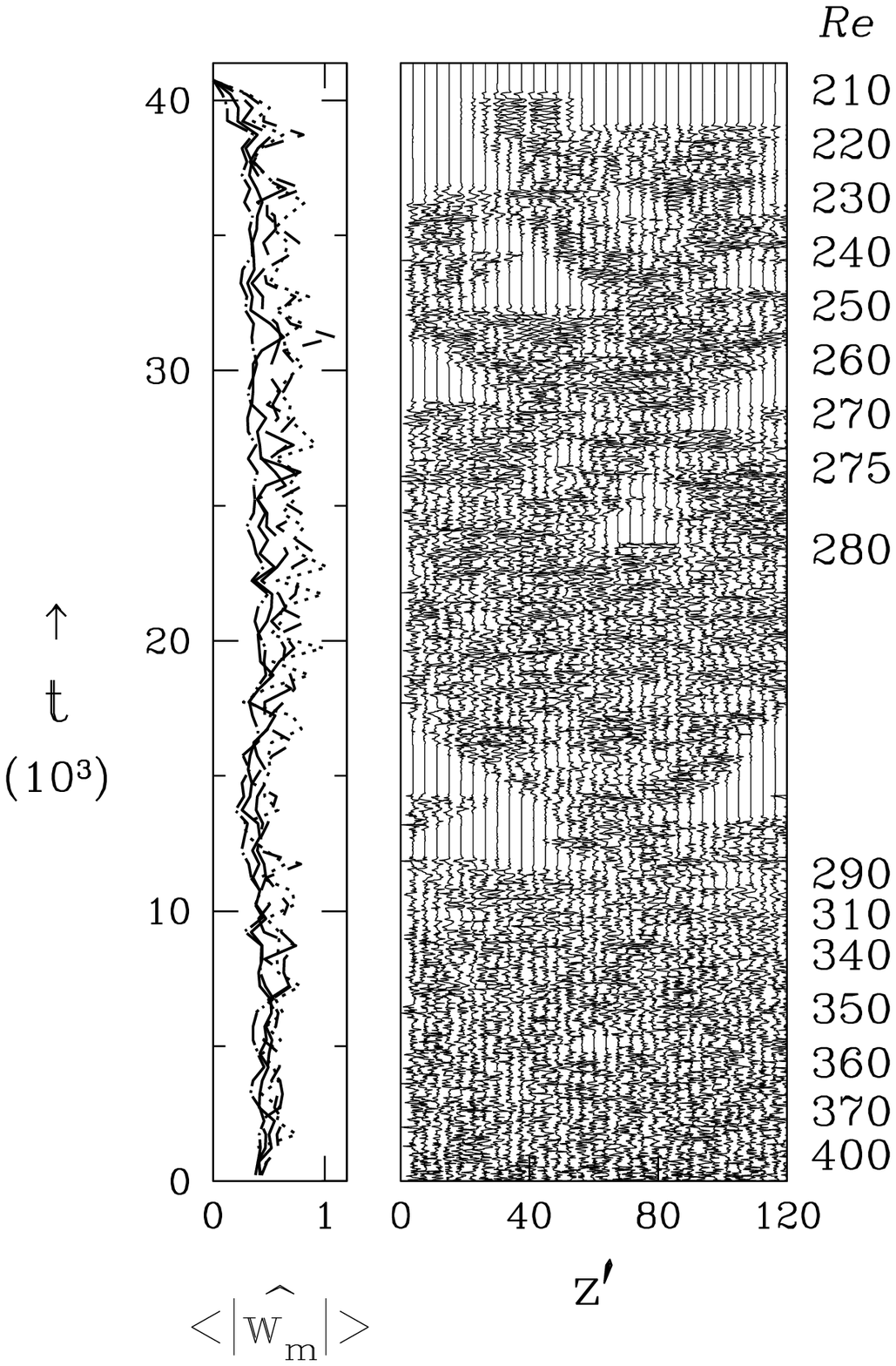}
\end{minipage}
\begin{minipage}{7cm}
\centerline{
\includegraphics[width=7cm]{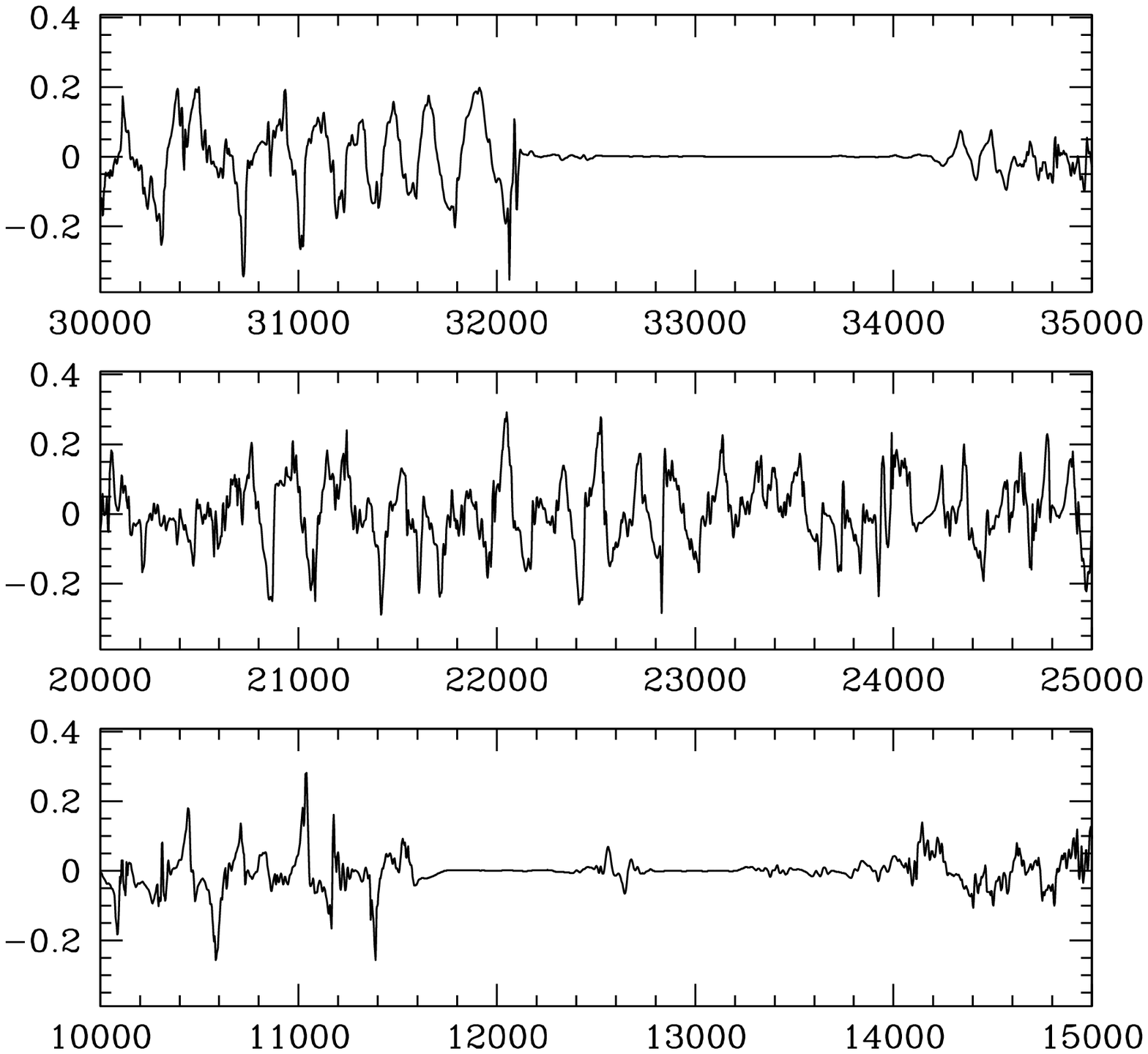}}
\end{minipage}
\caption{Evolution for $\theta=0^\circ$ with $L_\xp=L_x=10$ 
and $L_\zp=L_z=120$. $Re$ is lowered by steps of $\Delta Re = 5$;
not all intermediate values of $Re$ are shown.
The flow continues to have turbulent regions far below $Re=300$.
Left: evolution over the entire domain, showing the formation
and disappearance of turbulent domains.
Right: evolution of $w(x=0,y=0,z=60,t)$,
showing irregular periodic cycles.}
\label{fig:theta00_rip}
\end{figure}

\subsection{Long spanwise direction}

\noindent 
In the case $\theta=0^\circ$, the domain has a long spanwise direction 
($L_z = L_\zp=120$) and a
short streamwise direction ($L_x=L_\xp=10$).
In decreasing the Reynolds number by $\Delta Re = 5$ after each
interval of $T=1000$, we observe turbulent regions far below $Re=300$,
terminating only at $Re=210$, as shown in Figure~\ref{fig:theta00_rip}.  At several
times, the turbulence seems ready to disappear, only to spread out again.  In
order to confirm this surprising result, we have carried out longer
simulations at each of these low values of $Re$.  Turbulence persisted
over $T=4200$ (in the usual advective time units) 
at $Re=220$, over $T=3000$ at $Re=225$, and even
over $T=15000$ for $Re=230$.

Experiments \cite{Dauchot} and numerical simulations in large domains ($L_x \times L_y \times
L_z = 128 \times 2 \times 64$) \cite{Lundbladh} and numerical simulations in
periodic minimal flow units ($L_x \times L_y \times L_z = 4 \times 2 \times
6$) \cite{Hamilton} have produced long-lived turbulence only for $Re >
300$.  
A number of studies \cite{Schmiegel1,Schmiegel2,Faisst,EckhardtPROC} 
have examined turbulent
lifetimes as a function of initial perturbation amplitude, Reynolds number,
and quenching rate (rate of Reynolds number decrease) in minimal flow units.
In these studies, turbulence with a lifetime greater than $T=2000$ was
counted as sustained;
experiments \cite{Dauchot}, however, are carried out on timescales
several orders of magnitude longer than this.
\citeauthor{Schmiegel2} (\citeyear{Schmiegel2})
studied the effect of quenching rate on
turbulent lifetimes. For rates of Reynolds number decrease comparable
to ours, they found that turbulence could in some cases subsist to
$Re=280$ or 290 for times on the order of $T=1000$ to 10000;
for quenching rates ten times faster than ours, turbulence 
was occasionally sustained to $Re=240$.

If we compare our results to the previous simulations, 
then the conclusion would be that turbulence is favored
by a short streamwise direction $L_x=10$ and a long spanwise direction
$L_z=120$.  When either of these two conditions are lifted, the turbulence
disappears. We note, however, that 
our simulations do not systematically vary the initial conditions and
thus do not determine the probability of long-lived 
turbulence at these low Reynolds numbers near $220$.

We also note that Toh and colleagues \cite{TohJFM,TohPROC} have recently reported results
from simulations of Couette flow in domains with long spanwise extent compared
with the MFU geometry.  These simulations are for higher values of $Re$ than
those considered here.

We observe an approximately periodic oscillation in time, shown on the right
of Figure~\ref{fig:theta00_rip}.  The oscillation period of about 200 time
units has the same order of magnitude as the minimum turbulent cycle
\cite{Hamilton}, but further analysis of our results is
required before we can identify the streak and streamwise-vortex cycle
in our flow.

\section{SUMMARY}

\noindent We have used an extension of the minimal-flow-unit methodology to study
large-scale turbulent-laminar patterns formed in plane Couette flow.
Turbulent-laminar patterns are obtained as solutions to the Navier--Stokes
equations in domains with a single long direction.  The other
dimensions are just large enough to resolve the inter-plate distance and to
contain an integer number of lon\-gi\-tu\-di\-nal 
vortex pairs or streaks.  We have
presented various visualizations of the computed turbulent-laminar patterns as
well as space-time plots illustrating the formation and dynamics of these
patterns.  The time-averaged modulus of the spatial Fourier spectrum is shown
to provide a quantitative diagnosis of the patterns.  Periodic, localized, and
intermittent states occur in our simulations where similar states are observed
experimentally.

We have explored the patterns' dependence on Reynolds number, domain length
and tilt angle.  The patterned states do not appear to depend sensitively on
how the turbulence is initialized nor on the route taken to a particular point
in parameter space.  It is, however, possible that some parameter combinations
may support different numbers of turbulent bands (although we have not yet
observed this).  All states are bistable with respect to laminar Couette flow
and if parameters are changed too abruptly, then reversion to laminar Couette
flow occurs.

It appears that large-scale patterns are inevitable intermediate states on the
route from turbulent to laminar flow in large aspect-ratio Couette flow.  A
key open question is what mechanism causes laminar-turbulent patterns.  These
patterns are not only interesting in and of themselves, but may provide
clues to the transition to turbulence in plane Couette flow.  

\section*{ACKNOWLEDGEMENTS}

\noindent We thank Olivier Dauchot for valuable discussions and Ron Henderson for the
use of {\tt Prism}.  We thank the CNRS and the Royal Society for supporting
this work.  The two CPU decades of computer time used for this research were
provided by the IDRIS-CNRS supercomputing center under project 1119, and by
the University of Warwick Centre for Scientific Computing (with support from
JREI grant JR00WASTEQ).

\end{document}